\documentclass[aps,prl,twocolumn,showpacs,superscriptaddress]{revtex4}
\usepackage{graphicx}

\begin{document}
\title{High temperature expansion applied to fermions near Feshbach
resonance}
\author{Tin-Lun Ho}
\affiliation{Department of Physics,  The Ohio State University,
Columbus, Ohio 43210}
\author{Erich Mueller}
\affiliation{
Laboratory of Atomic and Solid State Physics, Cornell University, Ithaca, New York 14853
}
\date{revised: 02/16/04 EJM}

\begin{abstract}
We show that, apart from a difference in scale, all of the surprising recently observed properties of a {\em degenerate}
Fermi gas near a Feshbach resonance
persist in the high temperature Boltzmann regime.  In this regime,  the Feshbach resonance is unshifted.   By sweeping across the resonance,
a thermal distribution of bound states (molecules) can be reversibly generated.  Throughout this process,  the interaction energy is negative and continuous. We also show that this behavior must persist at lower temperatures unless there is a phase transition as the temperature is lowered.    We rigorously demonstrate universal behavior near the resonance.
\end{abstract}


\maketitle

At present, much experimental activity is concentrated on degenerate quantum gases near a Feshbach resonance \cite{
Duke,ENS,MIT,Jin,Grimm}
where the nominally weak effective interactions are strongly enhanced.
By applying a magnetic field, which moves the energy of a bound molecular state relative to the scattering continuum, the interactions can be tuned by by many orders of magnitude.
The scattering length diverges to negative/positive infinity when the molecular energy is infinitesmally above/below threshold.  At resonance, the scattering cross-section is limited only by {\em unitarity} ($\sigma=4\pi/k^2$, where $k$ is the relative momentum of the two atoms) and is {\em universal} (independent of the microscopic physics).
The universal nature of the unitarity limit leads to
a challenging many-body problem, since there are
no small parameters readily identifiable for the application of perturbation theory.

Recently, a sequence of beautiful experiments on Fermi gases have found unexpected and dramatic behavior near unitarity;
including a universal interaction energy \cite{Duke,ENS} and the reversible interconversion of atoms and molecules \cite{Jin}.  The former demonstrates the effects of unitary
scattering on bulk properties, and the latter suggests the possibility of an equilibrium phase with atoms and molecules in chemical equilibrium.
Several of the experimental results
are not well understood, and some even appear to be mutually contradictory \cite{ENS}.
In the following, we shall summarize these experiments and the fundamental questions which they raise.  We refer to the range of magnetic fields on either side of the resonance where the scattering length $a_{\rm sc}$ is positive or negative as the ``$+$'ve" and ``$-$'ve" side of the resonance.

\noindent {\bf (I)} {\em Universal interaction energy}:
Thomas's group at Duke \cite{Duke} has studied a gas containing two spin states of  fermionic
$^{6}$Li (which has a Feshbach resonance around 855 Gauss).
The experiment is performed on the negative side at 910 Gauss.
Upon release from the trap, the Fermi gas undergoes {\em anisotropic} expansion for temperatures between 0.1 and 3.5 $T_{F}$. This anisotropy can be explained by collisional hydrodynamics with a  {\em universal}
interaction energy proportional to the Fermi energy  ${\cal E}_{F}$.

\noindent {\bf (II)} {\em Properties near resonance:} More recently, Salomon's group directly measured the interaction energy of the same $^6$Li system, at temperatures between 0.5 and 1 $T_F$.
{\bf (a) } Crossing the resonance from the negative side, they find  that the interaction energy $\epsilon_{\rm int}$  remains  {\em negative and continuous}  across the resonance, despite the
expected
infinite jump of the scattering length. They find an interaction energy similar to that of the Duke group, convincingly demonstrating that $\epsilon_{\rm int}$ remains roughly constant over the temperature range  $0.1T_{F}$ to $T_{F}$.  {\bf (b)} Approaching the resonance from the opposite side, $\epsilon_{\rm int}$  is positive but drops to a negative value at a field of 700G; roughly 150G
before the resonance is reached.
{\bf (c)} The three-body recombination rate is maximal at 700 Gauss, rather than at 855 G; a result consistent with the work of Ketterle's group at MIT \cite{MIT}.
{\bf (d)} The anisotropy of the expansion (which is a measure of the interaction strength), is  maximal at 855 G.
While  {\bf (a)} through {\bf (c)} seem to indicate that the resonance is shifted from 855 G to 700 G  by many-body effects;  {\bf (d)} is consistent with an un-shifted resonance.

\noindent {\bf (III)} {\em Conversion between atoms and molecules:} Very recently, Jin's group at JILA has studied a gas consisting of two spin states of fermionic $^{40}$K \cite{Jin}. They show that molecules are produced as one crosses the resonance from the $-$'ve to the $+$'ve side.  Moreover, this process is reversible. This experiment, performed at $T=0.1T_{F}$ 
finds
no  significant shift of the original resonance.

Here, we focus on the Fermi gas near a Feshbach resonance.
Far from resonance, the interaction energy is
$\epsilon_{\rm int}=gn_{\uparrow}n_{\downarrow}$, where $g=4\pi\hbar^2 a_{\rm sc}/M$, $n_{\uparrow}$ and $n_{\downarrow}$ are the number densities of the two spin components. (We shall consider the case $n_{\uparrow}=n_{\downarrow}=n/2$.)
A key question is how this non-universal form of $\epsilon_{\rm int}$ turns into a universal function near resonance, (see {\bf (I)} and {\bf (II)}). Moreover, how does this function reflect the existence of molecules and what signatures do these molecules have if they exist?
Equally important is whether the original resonance is shifted in a many-body medium, (see {\bf (II)}).  To use methods where errors can be estimated precisely, we consider the
Boltzmann regime. In this case,  physical quantities can be calculated systematically through a  high temperature series expansion;  and yet the issues of the emergence of universal behaviors at unitarity still remain.

One might wonder whether the phenomena in the Boltzmann regime have any relevance to current ultralow temperature experiments.
The connection is simply that  {\em in the absence of any phase transition in the normal state},  the thermodynamic functions in the Boltzmann regime are analytically continuations (in $T$ and $n$) of those in the degenerate regime. This analytic continuation imposes strong constraints on the phase diagram, allowing one to qualitatively understand the physics at all temperatures.
In fact, apart from a difference in scale, the {\em exact} results in Boltzmann regime show
{\em all of the features} discovered in experiments ${\bf (I)}$, ${\bf (II)}$, and ${\bf (III)}$.
Surprisingly, the high temperature results, which should work well when $n_\uparrow\lambda^3=n_\downarrow\lambda^3=n\lambda^3/2\ll1$ (where $\lambda
=h/\sqrt{2\pi M k_{B}T}$ is the thermal wavelength and $k_B$ is Boltzmann's constant,) agrees reasonably well with the ENS experiments \cite{ENS} performed at $n \lambda^3/2=(4/3\sqrt{\pi})(T_F/T)^{3/2} \sim 1.6$ (see point ${\bf (E)}$, later).
In the following, we derive the interaction energy density in the Boltzmann regime, then draw a series of conclusions labeled below as ${\bf (A)}$ to ${\bf (D)}$.

{\bf Energy density in the Boltzmann regime:}   At high temperatures, or low densities,
the grand partition function ${\cal Z} = {\rm Tr}e^{-(H-\mu N)/k_{B}T}$ can be expanded in
the fugacity $z=e^{\mu/k_{B}T}$ \cite{Ul,LLSM}; and to the second order in $z$, where interaction effects first appear, is ${\cal Z}  = {\cal Z} ^{(o)} + 2\sqrt{2} \left( { V z^2}/{\lambda^3} \right)  b_{2}$,
where the superscript $``o"$ denotes quantities for non-interacting systems, and $b_{2}$$ =  $$\sum_{\nu}( e^{- \beta E_{\nu}^{(2)}} - e^{- \beta [E_{\nu}^{(2)}]^{(o)}} )$  is the second virial coefficient,
\begin{equation}
b_{2} = \sum_{b} e^{|E_{b}|/k_{B}T} + \sum_{\ell} \gamma_\ell\int^{\infty}_{0} \frac{ d k}{\pi}
\frac{ d\delta_{\ell}(k)}{dk} e^{ - \frac{\hbar^2 k^2}{mk_{B}T}} ,
\label{b2} \end{equation}
where $\gamma_\ell=2\ell+1$, the sum is over all integers $\ell=0,1,..$, $E_{b}$ is the energy of the two body bound state, and $\delta_{\ell}(k)$ is the phase shift of the $\ell$-th partial wave\cite{explainb2}.

A standard thermodynamic calculation \cite{Thermodynamics} shows that, to lowest nontrivial order in $n\lambda^3$, the energy density is
\begin{equation}\label{eps}
\epsilon = \frac{3nk_{B} T}{2} \left( 1  +  \frac{ n \lambda^3}{2^{7/2}} \right) +
 \epsilon_{\rm int} \equiv \epsilon_{\rm kin} + \epsilon_{\rm int}
\end{equation}
where $\epsilon_{\rm kin}$ and $\epsilon_{\rm int}$ are respectively the kinetic and interaction energy densities;
\begin{equation}
\epsilon_{\rm int} =  \frac{3k_{B} T n }{2} \left(   n \lambda^3 \right)
\left[ - \frac{b_{2}}{\sqrt{2}}  + \frac{\sqrt{2}}{3} T  \frac{\partial b_{2}}{\partial T} \right].
\label{key} \end{equation}
Since the contributions of the partial waves with $\ell\geq 1$ in eq. (\ref{b2}) are smaller than those of the s-wave by a factor $n\lambda^3$, we shall ignore them in our subsequent discussions.

Far from resonance the phase shift is $\delta(k) =  - a_{\rm sc} k$  for small $k$.
Due to the Gaussian cutoff in eq. (\ref{b2}), no errors are made by using this expression for all $k$.
In the absence of bound states, we then have $b_{2} = -a_{\rm sc}/(\sqrt{2} \lambda)$, and hence
$T\partial b_{2}/\partial T = b_{2}/2$. Equation (\ref{key}) then reduces to  the usual expression
$\epsilon_{\rm int} =  g n_{\uparrow}n_{\downarrow}$, with $n_{\uparrow}=n_{\downarrow}=n/2$.

{\bf Energy levels and interaction energy :}
To evaluate $\epsilon_{\rm int}$ near Feshbach resonance
 we use the expression for the phase shift  valid for $k<<b^{-1}$, where $b$ is the range of the potential\cite{LLQM}
\begin{equation}
k {\rm cot}\, \delta(k) = - \frac{1}{a_{sc}}  + \frac{r_{o}k^2}{2},
\label{cot} \end{equation}
where $r_{o} (\sim b)$ is the effective range and $a_{sc}$ is the s-wave scattering length. As a function of magnetic field, $a_{sc}$ varies as
$a_{sc} = a_{bg}\left(1 - \frac{\Delta B}{B-B_{o}}\right)$, where $B_{0}$ and $\Delta B$ are the location and the width of the resonance, and  $a_{bg}$ is the background scattering length away from resonance.  The region $B>B_{0}$ and $B<B_{0}$ where $a_{\rm sc}<0$ and $>0$ are referred to as the ``$-$'ve" and ``$+$'ve" side of the resonance. For temperature range such that thermal wavelength is larger than the range of the potential, $\lambda>b$, we can use eq.(\ref{cot}) in eq.(\ref{b2}) where  the integral is cutoff at $\lambda^{-1}$. Near resonance, $\lambda/a_{sc}, r_{o}/a_{sc} \rightarrow 0$, and
eqs.(\ref{b2}) and (\ref{cot}) imply that  up to terms of order $r_o/a_{sc}$,
\begin{equation}
b_{2} = \sum_{b}e^{|E_{b}|/k_{B}T} - \frac{{\rm sgn}(a_{sc})}{2}\left( 1 - {\rm erf}(x) \right)
e^{x^2},
\label{b22} \end{equation}
where $x= \lambda/(\sqrt{2\pi}a_{sc})$, and ${\rm erf}(x)$ is the error function.
We can then evaluate $\epsilon_{int}$ using eq.(\ref{key}) and (\ref{b22}). At the same time, it is useful to look at the effect of phase shifts on the energy levels.
In a box of size $R$, the wavevector $k$ is changed from its non-interacting value  $k_{o}=\ell  \pi/R$ ($\ell=1,2,3 ..$) to  $k  \sim k_{o}  -  \delta (k_{o}) /R$  through the boundary condition
${\rm sin}(kR+ \delta(k))=0$  at a large distance $R$. Scattering states and bound states correspond to real and  imaginary $k$ solution of this boundary condition.
The energies of  lowest few states in the center of mass frame ($E=\hbar^2k^2/M$) are shown in fig. 1.  As one passes through the resonance from the $-$'ve side, the lowest state in the continuum turns into a bound state, causing $\delta(k=0)$ to change abruptly from $0$ to $\pi$.  The corresponding interaction energy $\epsilon_{\rm int}$ calculated from eq.(\ref{key}) and eq.(\ref{b22}) are shown in fig. 2.
Despite the simplicity of the calculation, considerable information can be deduced:
\begin{figure}
\includegraphics[width=\columnwidth]{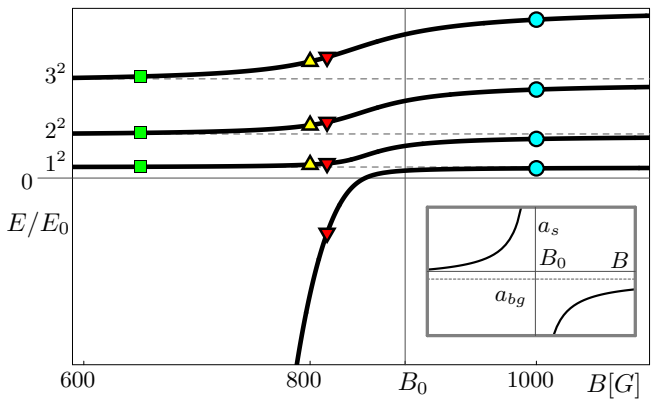}
\caption{(Color Online):
Energy levels of a two body system in the center of mass frame calculated from
the phase shift and the boundary condition in the text, with  $R=14\mu m$, $B_{0}= 855$G, $\Delta B = 325$G,  
and $a_{bg}= -120$nm.
The vertical scale is 
expanded to isolate individual energy levels.
The dotted lines are energy levels of a non-interacting system, $E_{n}^{(0)}=E_{0}n^2$, $E_{o}=\hbar^2 (\pi /R)^2 /M$, $n=1,2,3$. 
Passing from $B>B_0$ to $B<B_0$, the lowest state in the continuum becomes bound.
 Light (yellow) upright and dark (red) inverted triangles indicate states at the same field with and without the bound state occupied.
The interaction  energies due to thermal occupation of these states, and those marked by circles and squares, are shown in fig.2 by the same symbols.
The inset shows the behavior of the scattering length. }
\end{figure}
\begin{figure}[t]
\includegraphics[width=\columnwidth]{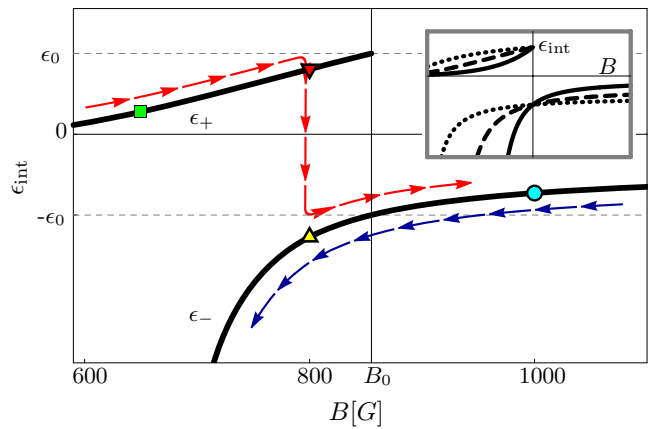}
\caption{
(Color Online): Interaction energy.
The energies $\epsilon^{(\pm)}_{\rm int}$   increase monotonically as the field is increased,
reaching the universal value $\pm\epsilon_{0}$ at resonance, where $\epsilon_{0} = (3nk_{B}T/2)(n\lambda^3/2^{3/2})$.  The negative branch, $\epsilon^{(-)}_{\rm int}$, is continuous across the resonance, while  $\epsilon^{(+)}_{\rm int}$ will jump to
 $\epsilon^{(-)}_{\rm int}$ if the bound state is occupied (say, at the field labeled by the triangle).
Inset shows the temperature dependence: solid, dashed and dotted lines correspond to $T=1,10,100\mu$K.}
\end{figure}

\noindent ${\bf (A)}$  Approaching the resonance from the $-$'ve side; $\epsilon_{\rm int}$ follows a ``negative branch"  $\epsilon_{\rm int}^{(-)}$  which is negative and decreases monotonically.  (See fig.2).
 $\epsilon_{\rm int}^{(-)}$ evolves
 from the (temperature independent) non-universal form $gn_{\uparrow}n_{\downarrow}$ far from resonance to a (temperature dependent) universal form $-\epsilon_0=-(3nk_{B}T/2)(n\lambda^3/2^{3/2})$ at resonance, and continues on to the $+$'ve side.  This universal form, which follows from the fact that $b_{2} = 1/2$ and $\partial b_{2}/\partial T =0$ at resonance $(x=0)$\cite{Ul}, is the high temperature analog of universal interaction found in {\bf (I)} and {\bf (II)}.

\noindent ${\bf (B)}$ Despite the change in sign of the scattering length, the interaction energy $\epsilon_{\rm int}^{(-)}$ remains negative across the resonance (shown in fig. 2) because of the thermal population $\langle n_{b} \rangle$ of bound states which exists when $a_{\rm sc}$ is positive,   
\begin{equation}
\langle n_{b} \rangle =   \frac{2\sqrt{2} z^2}{\lambda^3} e^{-E_{b}/k_{B}T} = n
\left( \frac{n\lambda^3}{\sqrt{2}} \right) e^{|E_{b}|/k_{B}T} + .. \,\,  .
\label{nb} \end{equation}
Although proportional to the small factor $n\lambda^3$, $\langle n_{b} \rangle$ is macroscopic in the thermodynamic limit.  Equation (\ref{nb}) in turn implies that {\em in a bulk system, the original resonance} (at field $B_{0}$) {\em can not be shifted to the positive side} (to  $B_{1}<B_{0}$) {\em at a lower temperature  unless there is a phase transition for all magnetic fields between $B_{1}$ and $B_{0}$ where $\langle n_{b}\rangle$  disappears as temperature is lowered from the Boltzman regime}.  (See fig. 3).  So far, such a phase transition has not been observed.
Should future experiments rule out such a transition in $^{6}$Li,  one must then conclude that
the resonance in the ENS experiment\cite{ENS} is not shifted.   The absence of a shift of the resonance is also consistent with the findings in ref. \cite{Jin} for $^{40}$K.

\begin{figure}[t]
\includegraphics[width=\columnwidth]{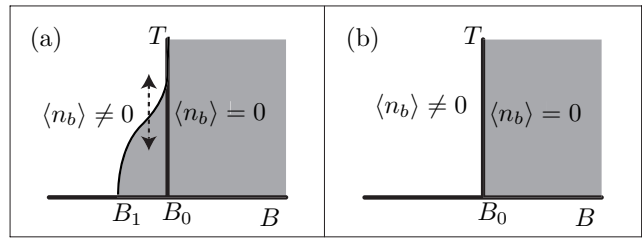}
\caption{ Schematic phase diagrams.  Because the original resonance $(B_{0})$ is un-shifted at high temperature, a shift in resonance at low temperature $(B_{1}<B_{0})$ will imply a phase boundary as shown in (a), which means that  $\langle n_{b}\rangle$ will disappear (or appear) as temperature is
lowered (or raised) along the path indicated by the double-headed arrow.  The phase diagram for an unshifted resonance is shown in (b). }
\end{figure}

\noindent ${\bf (C)}$ Approaching the resonance from the $+$'ve side;  if the bound states are not occupied, $\epsilon_{\rm int}$ will follow a ``positive branch" $\epsilon_{\rm int}^{(+)}>0$ which increases monotonically, evolving from $gn_{\uparrow}n_{\downarrow}$ to $\epsilon_0=(3nk_{B}T/2)(n\lambda^3/2^{3/2})$ at resonance (see fig. 2). However,  a gas containing only scattering states is not in true equilibrium, as the latter requires that all states, including bound states, are thermally populated.
As pointed out by Petrov \cite{petrov}, the three-body collisions which convert scattering states into bound states can only lead to chemical equilibrium if the released energy is insufficient to eject the particles from the trap.  If one is sufficiently far from the resonance, then the molecular binding energy is larger than the trap depth, and the three-body collisions instead lead to loss.  The sudden drop in the interaction energy near 700G in the ENS experiment \cite{ENS} is consistent with the production of trapped molecules.  Moreover, Jochim et al. \cite{Grimm} have directly observed the equilibration of atom/molecule populations near this field.
This scenario also predicts that the development of an equilibrium molecular population will coincide with a peak in the three body loss rate.  This peak is observed at ENS \cite{ENS}, MIT \cite{MIT}, and Innsbruck \cite{Grimm}.

\noindent ${\bf (D)}$ The extension of $\epsilon_{\rm int}^{(-)}$ to the positive side of the resonance is due to the  population of the bound state which becomes available there. If the system is brought across the resonance adiabatically,  quasi equilibrium is maintained and the
process is reversible. Consequently,  molecules generated on the positive side will turn back to atoms when the system is brought back to the negative side. This is consistent with the experiments on $^{40}$K  \cite{Jin}.

\noindent ${\bf (E)}$ The experiment at ENS was performed at temperature $T=3.5\mu$K and degeneracy factor $T/T_{F}=0.6$, corresponding to $n\lambda^3/2=1.6$.  Using the same temperature, we have plotted the ratio
$\epsilon_{\rm int}/\epsilon_{\rm kin}$ in fig. 4 for $T/T_{F}= 1.2$, 0.6, and 0.4, corresponding to $n\lambda^3/2 =0.6, 1.6, 3.0$.  
Although our result on interaction energy should only be accurate when $n\lambda^3 /2<1$, we
also plot it at higher phase space densities (i.e. extending it to regions $n\lambda^3 /2>1$) to indicate its temperature dependence. Moreover, these extensions are the leading term of the interaction energy in powers of $n\lambda^3 /2$. 
On the positive side all three of our curves are consistent with the experimental data, while on the negative side the $T/T_F=1.2$ curve is closest.
The fact that the $B>B_0$ data matches this higher temperature curve
may be due to higher order terms in the high temperature expansion or systematic differences in the density/temperature of the sample on the two sides of the resonance.  
In any case, it is clear is that a Fermi gas in the Boltzmann regime exhibits all the phenomena seen in current experiments. Equally important is the fact that the exact high temperature results near unitarity forces one to conclude the existence of a phase where atoms and molecules are in chemical equilibrium.

Having shown that the behavior of the quantum gas in Boltzmann regime has all the
characteristics seen in experiments at lower temperatures, we reiterate that the two regimes are connected
by the analyticity of the thermodynamic functions.  Despite the divergent scattering length, at small fugacity $z$ (or equivalently small density $n\lambda^3$) the probablility of particle collisions is small, leading to our systematic perturbative scheme.

We thank John Thomas and Christophe Salomon for stimulating discussions.
This work is inspired by Professor Lev Pitaevskii's talk on the virial coefficients during his 70th birthday celebration which led us to think about  the recent experiments.
This work is supported by NASA GRANT-NAG8-1765  and NSF Grant DMR-0109255.

\begin{figure}[b]
\includegraphics[width=\columnwidth]{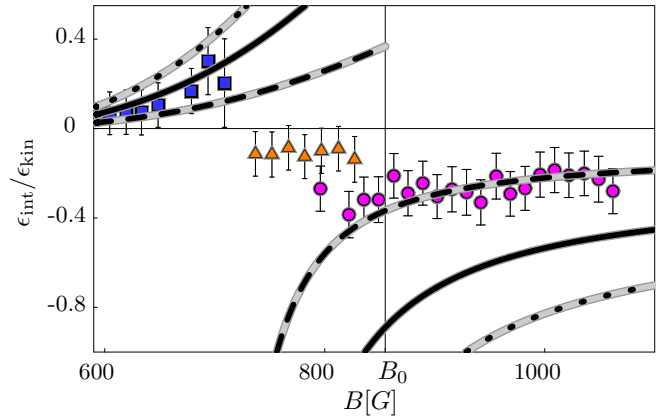}
\caption{(Color Online):  The ratio $\epsilon_{\rm int}/\epsilon_{\rm kin}$ for $T=3.5\mu$K.  Squares (circles) show data 
from ref. \cite{ENS}
which agrees with $\epsilon^{+}$, ($\epsilon^{-}$). Triangles do not fall on either curve and probably reflect a nonequilibrium situation.
The dashed, solid, and dotted lines are for $T/T_{F}=1.2, 0.6, 0.4$ respectively. }
\end{figure}


\begin{thebibliography}{99}
\bibitem{explainb2} Eq.(\ref{b2}) is obtained by noting that the wavevector $k$ is changed by the phase shift from the non-interacting value $k_{o}$ which leads to change of density state given by $d\delta(k)/dk$. The details are given in \cite{LLSM}.
\bibitem{Thermodynamics}
Using the relation $PV =k_{B}T{\rm ln} {\cal Z}$, the pressure 
 is
$P(T, \mu)$$=$$P^{(o)}(T,\mu)$$+$$2\sqrt{2}\left({ k_{B} T z^2}/{\lambda^3} \right)b_{2}$,~where $P^{(o)}(T,\mu)$ is the pressure for a two component non-interacting Fermi gas.  The fugacity expansion for $P^{(o)}(T,\mu)$ (see Ch. 5  in \cite{LLQM}) gives
$P(T,\mu)$$=$$2\lambda^{-3} k_{B}T$$ \left(z-2^{-5/2}z^2 +  \sqrt{2} b_{2} z^2\right)$.
Using the Gibbs-Duham relation $dP=nd \mu+s d T$, where $n$ is the number density and $s$ is the entropy density, the density is
$n(T, \mu)$$=$$2 \lambda^{-3}$$\left( z - 2^{-3/2}z^2  +   2\sqrt{2}  b_{2} z^2 + ..\right) ,$ which is inverted to produce
$z$$=$$(n\lambda^3/2)$$+$$\left(  2^{-3/2} -2\sqrt{2} b_{2} \right)$
$\times\left(n\lambda^3/2\right)^2$$+\ldots$
Similarly, it is straightforward to show that
$s =   5P/(2T)  - \mu n/T +
2\sqrt{2} k_{B} T z^2 \lambda^{-3}   \partial b_{2}/\partial T$.
The energy density,
$\epsilon = Ts + \mu n - P$, then becomes
$\epsilon  =  ({ 3k_{B} T z}/{\lambda^3} )
   [  1 $$+ z (   {-2^{-5/2}}
 $$+ \sqrt{2} b_{2}$$+ ({2\sqrt{2}}/{3})  T  $$({\partial b_{2}}/{\partial T}) )   ]$,
 which can be written in the form of Eq.~(\ref{eps}).
\end{thebibliography}
\end{document}